# Donor-acceptor recombination emission in hydrogen-terminated nanodiamond: Novel single-photon source for room-temperature quantum photonics

D. G. Pasternak[1*]; A. M. Romshin[1]; R.H. Bagramov[2]; A.I. Galimov[3]; A.A. Toropov[3]; D. A. Kalashnikov[4]; V. Leong[4]; A.M. Satanin[5]; O. S. Kudryavtsev[1]; A.L. Chernev[6]; V.P. Filonenko[2]; I. I. Vlasov[1*]

[1]Prokhorov General Physics Institute of the Russian Academy of Sciences, 38 Vavilova str, Moscow 119991, Russia.

[2]Vereshchagin Institute of High-Pressure Physics RAS, 14 Kaluzhskoe shosse, Moscow, Troitsk 108840, Russia.

[3]Ioffe Institute, 26 Politekhnicheskaya str, 194021, St. Petersburg, Russia.

[4]Institute of Materials Research and Engineering (IMRE), Agency for Science Technology and Research (A*STAR), 2 Fusionopolis Way, Innovis #08-03, Singapore 138634, Republic of Singapore.

[5]National Research University Higher School of Economics, 20 Myasnitskaya str, Moscow 101000, Russia.

[6]Laboratory of Nanoscale Biology, Institute of Bioengineering, Ecole Polytechnique Federale de Lausanne, 1015, Lausanne, Switzerland.

**Abstract**

*In fluorescence spectra of nanodiamonds (NDs) synthesized at high pressure from adamantane and other organic compounds, very narrow (~1 nm) lines of unknown origin are observed in a wide spectroscopic range from ~500 to 800 nm. Here, we propose and experimentally substantiate the hypothesis that these mysterious lines arise from radiative recombination of donor-acceptor pairs (DAPs). To confirm our hypothesis, we study the fluorescence spectra of undoped and nitrogen-doped NDs of different sizes, before and after thermal oxidation of their surface. The results obtained with a high degree of confidence allowed us to conclude that the DAPs are formed through the interaction of donor-like substitutional nitrogen present in the diamond lattice, and a 2D layer of acceptors resulting from the transfer doping effect on the surface of hydrogen-terminated NDs. A specific behavior of the DAP-induced lines was discovered in the temperature range of 100-10 K: their energy increases and most lines are split into 2 or more components with decreasing temperature. It is shown that the majority of the studied DAP emitters are sources of single photons, with an emission rate of up to >1 million counts/s at room temperature, which significantly surpasses that of nitrogen-vacancy and silicon-vacancy centers under the same detection conditions. Despite an observed temporal instability in the emission, the DAP emitters of H-terminated NDs represent a powerful room-temperature single-photon source for quantum optical technologies.*

**Introduction**

Since the beginning of the 2000s, diamond containing fluorescent point defects have been considered as one of the most promising material platforms for creating solid-state single photon emitters (SPEs) [1]. Single-photon emission from various diamond defects covers the wide visible-NIR range. Diamond SPEs are promising for the development of protected free-space optical communication lines [2] and other applications of quantum-optical technologies of the 21st century. The main advantages of diamond SPEs are the possibility of their use at room temperature (RT), narrowband emission (<10 nm), infinite photostability. Until now, diamond SPEs have been based on intra-center electron transitions.

A new generation of nanodiamonds (NDs) produced from various hydrocarbons at high pressure and high temperature (HPHT) [3-5] combines the high structural quality of the diamond lattice with an extremely high concentration of surface terminating hydrogen [6,7], which, in turn, stimulates the formation of quasi-2D surface layer of free carriers (holes) [8]. Optical studies of this nanomaterial reveal to us still unexplored facets of the unique diamond properties. Thus, we recently discovered destructive phonon-plasmon interference, which manifests itself in the form of a narrow peak of transparency in the infrared absorption of such nanodiamonds [7]. Even earlier in the fluorescence spectra of undoped HPHT nanodiamonds synthesized from adamantane, we found an unusual picture - a "palisade" of narrow lines extended over a wide visible range [9]. Recently, A. Y. Neliubov et al. [10] observed similar lines in micron-sized HPHT diamonds in 600-650 nm range and showed that these unusual lines originate from single photon emitters. However, the origin of these emitters still remained a mystery.

The presence of narrow lines in a wide spectral range is typical for radiative recombination of donor-acceptor pairs (DAP) in semiconductors [11-15]. For example, sharp fluorescent lines related to DAP formed by nitrogen and boron impurities have been observed in diamond within the energy range 2.0—2.7 eV at temperatures below 200 K [12]. Recently, DAP-related fluorescence has been theoretically predicted [14] and experimentally observed in h-BN within 1.6-2.8 eV range at low temperatures [15].

We hypothesized that the new lines observed in hydrogenated HPHT diamonds may also belong to laser-induced DAP recombination. The most likely candidate for the role of donor in our case may be nitrogen, the most common impurity in diamond. The presence of nitrogen in undoped HPHT ND can be indicated by weak lines of NV centers in fluorescence spectra [16]. The most common acceptor impurity in diamond is boron, but we exclude its participation in the DAP of hydrogenated NDs, since we did not observe any luminescence at room temperatures in boron-doped nanodiamonds synthesized from organic compounds [17, 18]. It is known that transfer doping effect in hydrogen-terminated diamond induces the formation of surface acceptors (presumably related to atmospheric adsorbate) [6]. In our opinion, namely these acceptors most likely participate in the DAP recombination generating narrow-line fluorescence.

To verify our hypothesis about the link between narrow fluorescent lines (NFLs) in hydrogenated nanodiamonds and radiative recombination between nitrogen donor from diamond volume and surface acceptors, in this work we studied the fluorescence spectra of diamond nanoparticles (1) of different sizes, (2) doped with nitrogen, (3) after thermal oxidation of their surface. We also investigated the effect of temperature on the main spectral characteristics of individual NFLs in the range of 300-10 K.

## Results and discussion

*Size effect*

To check the relation of the NFL source to a nanodiamond surface, the fluorescence spectra of undoped diamond nanoparticles with a size of 100 nm - 1 μm were studied. NFLs were detected in a wide range of wavelengths from 500 (2.48 eV) to 800 (1.54 eV) nm. Most of the lines are observed upon laser excitation of clusters consisting of smallest nanoparticles of 100-300 nm in size (Fig. 1a). Position of lines and their number are not preserved in spectra when moving from one ND cluster to another. In the spectra of large individual crystallites (0.5-1 μm) NFLs are either not observed at all or are barely distinguishable (Fig. 1b). We will discuss the reasons for this later, in the last section of the "Results and Discussion". A significant increase of the NFL amount with increasing surface/volume ratio in the probed NDs is the first evidence of a connection between the source of new lines and the ND surface.

*Nitrogen doping*

To verify the connection of NFL source with nitrogen impurity in the diamond volume, the fluorescence spectra of nitrogen-doped NDs were studied. Nitrogen is added to the precursor in a ratio N/C ~ 1/1000 at., as part of adamantane carbonitrile (ACN) or detonation NDs (see Methods). In both cases, the samples show similar fluorescence spectra, which are dominated by the band of H3 (or N-V-N) center characterized by a zero-phonon line (ZPL) at 503 nm and a phonon sideband maximum at 520 nm (Fig 1c,d). However, in the spectra of samples obtained with the ACN, quite intense bands from $NV^o$ and $NV^-$ centers with narrow ZPLs at 575 nm and 738 nm, respectively, are also observed in the region 575-750 nm. The fluorescence of all 3 centers overlaps with NFLs, which makes their individual characterization difficult. Taking into account the lesser influence of the fluorescent background from $NV^o$ and $NV^-$ centers in the samples obtained with the detonation NDs, we carried out further characterization of narrow lines specifically for these kind of samples. The number of NFLs in the spectra of 100-300 nm N-doped diamonds increases significantly compared to the spectra of undoped NDs of the same sizes. The scatter in intensity between the new lines is significant (exceeds an order of magnitude). In addition, over time, the most of the lines, but not all, blink independently of each other, as we have shown in [9]. Statistical pattern of the NFL distribution over wavelengths, based on 150 lines recorded for different diamond clusters of the same sample, is shown in Figure 1e. Two maxima are found in this distribution, at 540 nm and 720 nm. With a further decrease in ND sizes (achieved by decreasing the synthesis temperature), we obtain an even denser arrangement of NFLs in one spectrum. Thus, for the clusters of diamond particles < 30-nm individual lines begin to overlap, forming two broad bands with maxima at 540 nm and 720 nm (Figure 1f), the same maxima as in Figure 1e. Note that brightest NFLs not depend on ND size, they are comparable in intensity to the Raman peak of diamond, both for <30 nm and 100-300 nm particles.

The significant increase in the number of NFL lines found for N-doped nanodiamonds compared to undoped ones confirms the connection between the source of the lines and the nitrogen impurity in the diamond volume. For comparison, we have not established the influence of other dopants (for example, silicon and germanium) on the characteristic NFL distribution in a spectral range of 500-800 nm.

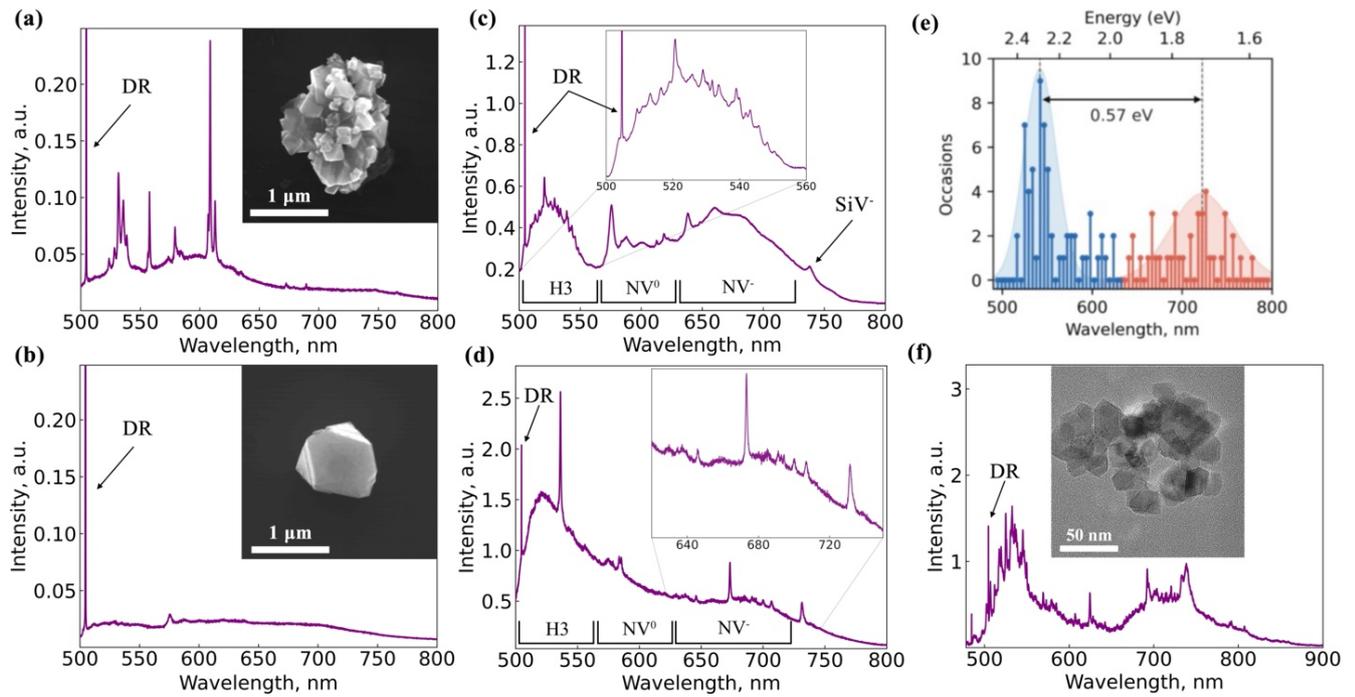

*Figure 1. Photo-luminescence (PL) spectra (a) of 100 - 300 nm «undoped» NDs (inset: the SEM image of typical ND cluster used for spectroscopy); (b) of «undoped» individual diamond ~1 um (inset: the SEM image of typical diamond crystallite used for spectroscopy); (c) of 100-300 nm NDs doped with nitrogen from ACN (inset: a zoomed spectrum part containing numerous NFLs). Observed H3, NV⁰, NV⁻ bands and SiV⁻ peak are properly marked; (d) of 100 - 300 nm NDs doped with nitrogen from DND (inset: a zoomed spectrum part containing numerous NFLs). (e) Statistics of NFL position distribution for N-doped 100 - 300 nm NDs. (f) PL spectrum of N-doped NDs <30 nm (inset: the TEM image of typical NDs used for spectroscopy). Two maxima are observed at 540 nm and 720 nm in (e) and (f). All PL spectra were recorded under 473-nm (2,62 eV) laser excitation at room temperature (RT) and normalized to the intensity of the diamond Raman (DR) peak at 505 nm.*

*Annealing*

To check the connection of the NFL source with the acceptors resulting from the transfer doping effect on the surface of H-terminated nanodiamonds, we decreased the hydrogen concentration by thermal oxidation of the ND surface in air. As we have previously established, for HPHT diamond crystallites with a size of more than 100 nm, heating to 600-650 C is required for effective removing hydrogen from the diamond surface [9]. PL spectra of 100-300 nm N-doped NDs recorded before and after air heating at 620 ºC are shown in Figure 2. As we see, most of the NFLs disappear after annealing. In particular, in the range of 620-800 nm all NFLs disappear and only lines associated with NV⁻ and SiV⁻ remain (Fig.2, insert). The results obtained indicate the connection of NFLs with the surface acceptors in hydrogen-terminated NDs.

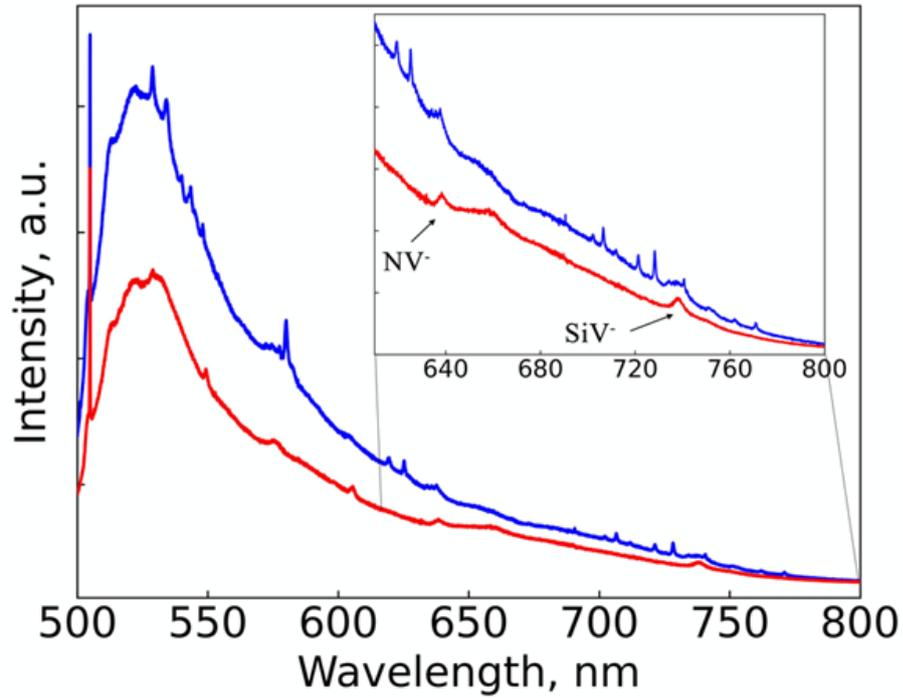

*Figure 2. PL spectra of 100-300 nm N-doped NDs recorded before (blue) and after (red) air heating at 620 °C. All PL spectra were recorded under 473-nm (2,62 eV) laser excitation at RT. in the range of 620-800 nm all NFLs disappear and only lines associated with NV⁻ and SiV⁻ remain (insert).*

*NFL characterization in the temperature range 10-300 K*

The main spectral characteristics of individual NFLs were studied at room and cryogenic (10 K) T, as well as their temperature dependences in the range of 10-300 K. The studies were carried out on clusters of 100-300 nm diamond crystallites doped with nitrogen, in the spectra of which, unlike smaller crystallites, predominantly discrete NFLs are observed. It has been established that the characteristic width of the lines, determined as a full width at half-maximum (FWHM) of their intensity, lies in the range of 0.5-2 nm (1.5-6 meV) (Fig. 3a). The second-order correlation functions ($g^{(2)}$) for single NFLs were measured under 660 nm (1.88 eV) laser excitation in order to exclude the influence of fluorescence of nitrogen-related centers. It was found that for the most lines in the range of 660-800 nm (1.88 – 1.55 eV) $g^{(2)}(0) < 0.5$ indicating their relation to single photon emitters (Fig. 3b). The maximum intensity of a single fluorescent line recorded in our experiments was 1.2 million counts/s at laser power 1.7 mW (Fig. 3c). For a number of lines, fluorescence lifetimes were determined at a level of 1 - 2 ns.

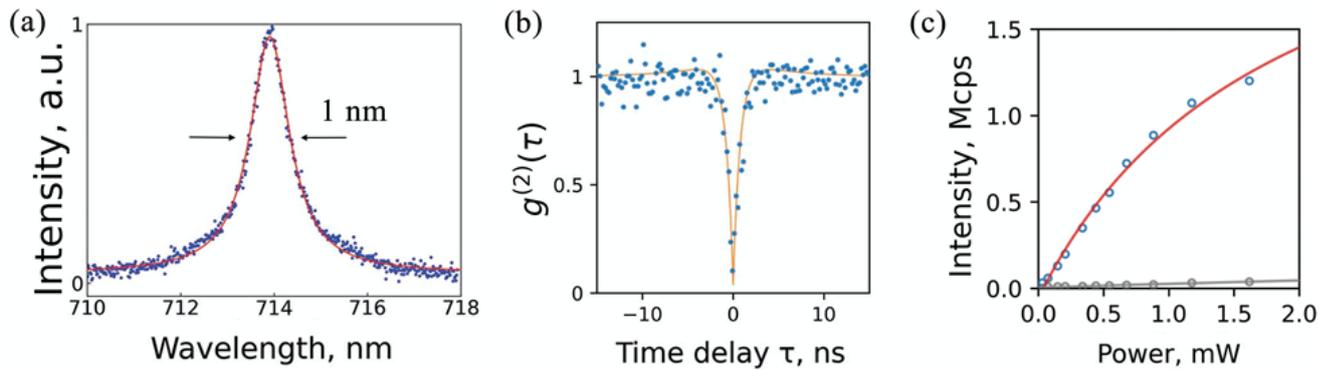

*Figure 3. Room-temperature characterization of typical single NFL in PL spectra of 100-300 nm N-doped NDs. The spectra were recorded under 660-nm (1,88 eV) laser excitation. (a) The NFL with FWHM = 1 nm positioned at 714 nm. Experimental data (dots) are well fitted with Lorentzian profile (red curve). (b) Second-order autocorrelation function with $g^{(2)}(0)$ <0.1 was measured for that NFL. The orange curve is a best-fit to the experimental data (dots). (c) The dependence of the NFL intensities (circles) on excitation laser power (fitted with red curve). Typical background fluorescence from silicon substrate is fitted with grey curve.*

As T decreases to 10 K, the number of lines increases several times, they become narrower and more intense (Fig. 4a). Narrowband fluorescence becomes more stable in time. Most of the studied lines show a pronounced polarization dependence (Figure 4b), which, along with the $g^2$ measurements for some of these lines, indicates their correspondence to single photon sources. Fluorescence lifetimes determined for some lines at 10 K show virtually the same values (1–2 ns) as at room T.

Let us turn to the results on the temperature dependences of the spectral width and position of the NFLs. For all studied lines, a gradual decrease in their width is observed from ≈1 nm (3 meV) at RT to ≈ 40 pm (0.2 meV) at 100 K (Figure 4c). Some of the lines remain single with decreasing T (Fig. 4d), whereas most of the them exhibit a doublet and triplet structure (Fig. 4e). The observation of possible further decrease in a linewidth is limited by the 40 pm spectral resolution of the fluorescence detection system used. When using the scanning resonant excitation of NFL with registration of its possible phonon sideband, shifted relative to the line into the long-wave region, the line was not detected. This technique provides a spectral resolution of 5 MHz and was successfully used by us to detect narrow lines of single SiV centers at 10 K [19]. The disappearance of NFLs upon their resonant excitation means an extremely weak coupling between the excited DAP and lattice phonons. The temperature dependence of the NFL positions behaves somewhat unusually. Down to a T of about 90 K, the lines shift toward shorter wavelengths, that is typical for thermally broadened two-level transitions; however, with a further decrease in T, the lines slightly shift by ≈50 pm in the opposite direction. As a hypothesis, we explain the appearance of doublet, triplet, and possibly more complex NFL structures by the interaction of closely located surface acceptors of two or more donor-acceptor pairs, which leads to the splitting of the energy levels of each DAP. The coincidence of the polarization dependences for the split lines indicates that these lines belong to the same center, although this claim requires additional confirmation. The unusual shift of lines to the red region at low temperatures may be associated with a change in the near-surface band bending with T.

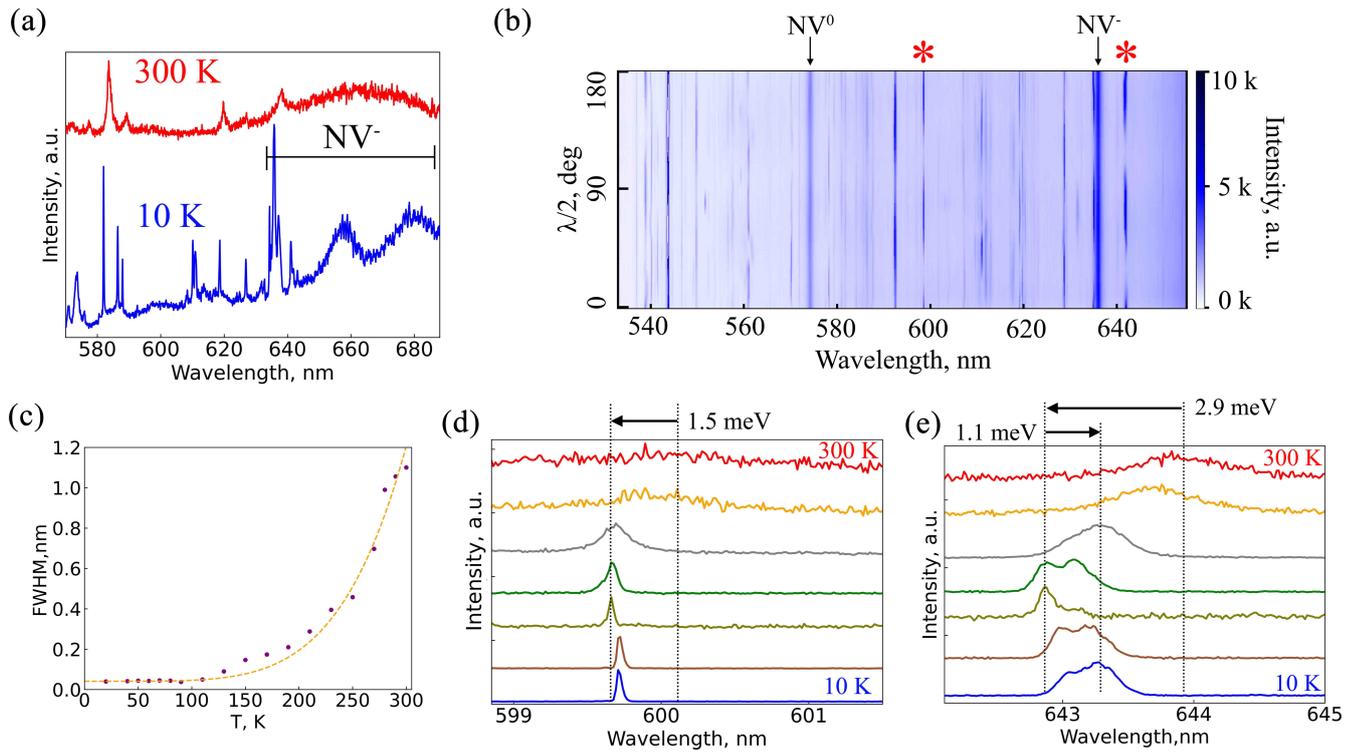

*Figure 4. Temperature-dependent characterization of NFLs in PL spectra of 100-300 nm N-doped NDs. The spectra were recorded under 532-nm (2.33 eV) laser excitation in a temperature range of 10-300 K. (a) PL spectrum recorded in 570-690 nm rage at 300 K (red) and 10 K (blue) (b) Polarization dependence measurements at 10 K in 535- 670 nm range. ZPLs of $NV^0$ and $NV^-$ (marked with black arrows) are not sensitive to changes in polarization angle, whereas the most NFLs demonstrate high polarization sensitivity. The NFLs marked with red asterisks show the highest polarization sensitivity. (c) Temperature dependence of the NFL at 600 nm (300 K). Dots are experimental data. The dotted orange curve demonstrates $(aT^3+b)$ dependence of the line width on T. (d) Temperature-dependent PL spectrum of the 600-nm NFL. (e) Temperature-dependent PL spectrum of 644-nm NFL. In (d) and (e) the PL spectra were recorded at 10, 50, 90, 130, 180, 270 and 300 K (from the bottom to top).*

### *Donor-acceptor pairs in H-terminated nanodiamonds*

The results of studying the dependence of the NFL amount in the fluorescence spectra on the N-doping of NDs, the diamond size, and the functional state of the ND surface allow us to assert that the new lines are associated with both nitrogen centers in the diamond volume and with its surface. Based on the established connection, as a model of the NFL source, we consider a donor-acceptor pair consisting of a single substituting nitrogen atom (N), as a donor, and some acceptor (A) appearing on the surface of hydrogenated ND. The presence of such interacting pairs was confirmed by Ristein et al. [20, 21] using p-type conductivity measurements on the surface of nitrogen-doped and hydrogen-terminated diamonds. It was found there that nitrogen located near a diamond surface effectively compensates for surface acceptors, and the resulting space charge induces a band bending that pulls the valence band up to the common Fermi level of the system (Figure 5a). Pairs consisting of interacting donors and acceptors create a set of electronic levels in the semiconductor band gap. When these levels are excited by laser, a series of sharp fluorescence lines in a broadband wavelength range appears [22, 23]. The energy of photons ($E_{DAP}$) emitted as a result of DAP recombination depends on the difference between the energy levels of the donor and acceptor ($E_D$-$E_A$) and the Coulomb component of their interaction in the charged state $e^2/\mathcal{E}R$ (where e is the electronic charge, $\mathcal{E}$ is the static dielectric constant, R is the distance between acceptor and donor), e.i. is a function of the relative inverse pair separation [23]:

$$E_{DAP} = (E_D - E_A) + \frac{e^2}{\mathcal{E}R} \qquad (1)$$

If the donor and acceptor centers are located at the sites of the diamond lattice, then in the spectrum of their recombination radiation the lines are observed at certain wavelengths, as ($E_D$-$E_D$) is constant in that case and the pair separation R assumes discrete values, proportional to the C-C bond length. As a result, observed spectrum are only slightly sample dependent [12]. In our case, the energy difference ($E_D$-$E_A$) near the diamond surface becomes a variable value depending on various factors. This difference will depend on the distance of the nitrogen donor to the surface, and on the magnitude of the band bending, which, in turn, depends on the donor, acceptor, and hydrogen concentrations. As a result, the NFL spectral distributions for different ND clusters can differ significantly within the same sample, exactly what is observed in our experiments. Nevertheless, in the statistical distribution of lines over the spectrum obtained for 100-300 nm diamond clusters (Fig. 1e), as well as in the fluorescence spectra of smaller NDs (Fig. 1f), we find two maxima with energies of 2.3 eV and 1.7 eV. This may be due, for example, to the two most characteristic values of band bending, corresponding to two predominant crystallographic orientations of the diamond surfaces, or to two types of acceptors involved in recombination with nitrogen and having different energy levels.

The rare detection of NFLs in large single crystals can also be explained by the DAP dipole direction perpendicular to the diamond surface. Indeed, at small angles of incidence of exciting radiation on the diamond surface, such dipoles will be excited ineffectively, since the radiation polarization will be near perpendicular to the dipole axis. In a variety of diamond crystal configurations, angles between adjacent facets typically exceed 115 degrees. Thus, in the cubo-ctahedron shape characteristic of diamond, the dihedral angle is 125 degrees. In such crystallites, the excited dipole radiation propagating along the initial facet will undergo multiple total internal reflection from subsequent facets. Thus, even with efficient dipole excitation, the radiation collection efficiency will be low for large crystallites (Fig. 1b). In NDs with a size of <30 nm, which is significantly smaller than the laser excitation wavelengths (~500 nm), there is no total internal reflection, which, along with a significant increase in the area/volume ratio, leads to a sharp increase in the NFL intensity and amount (Fig. 1f).

The DAP emission corresponding to individual lines demonstrates a pronounced polarization dependence, i.e. DAPs are dipole type emitters. The location of acceptors exclusively on the surface of the diamond particle, and donors in its volume, determines the specific distribution of dipole moments - all of them are oriented towards the diamond faces (Fig. 4b).

The qualitative picture of the electronic configuration of donor-acceptor pairs seems to us to be as follows. Let us consider a single donor nitrogen center with a Bohr radius of ~1 nm [12] and acceptor centers with an uncertain Bohr radius, localized in a 2D layer on the diamond surface. It is known that the optical matrix element responsible for the dipole radiation is determined by the matrix elements of the dipole moment (DM). The transverse (XY) component of the DM, lying in the plane of the surface, is proportional to the overlap factor of the DAP wave functions along the Z axis perpendicular to the surface, while the contribution to the longitudinal (Z) component of the DM contains an additional factor proportional to the projection of the dipole moment operator along the Z direction. It follows that if the characteristic distance (R) between the donor and the acceptor layer is comparable to the size of the acceptor wave function along the Z direction, the dipole moments will have noticeable components in the XY directions, which can lead to a significant broadening of the levels of individual dipole transitions. If the typical R well exceeds the size of the acceptor wave function, then the orientation of the dipoles perpendicular to the surface will predominate. We believe that it is the second case that occurs in the NDs under study, since we observe unusually narrow lines of single dipole emitters for room temperatures, which vary slightly in width in their large ensemble.

We do not exclude that similar DAP fluorescence was observed previously in H-terminated diamonds of various origins, but the observed NFLs either remained unexplained or were attributed to energy-shifted lines of already known color centers in diamond. For example, work [24] reported the detection of NFLs of an unknown nature in nanodiamond clusters synthesized by the CVD method and containing a nitrogen impurity. The lines were observed in the visible and near-IR regions, their width was 1–2 nm at room temperature, and the lines narrowed to 50 pm at T = 10 K. Since the surface of the CVD ND is terminated with hydrogen, we believe that the sources of those NFLs observed is the same pairs of nitrogen donors and surface acceptors.

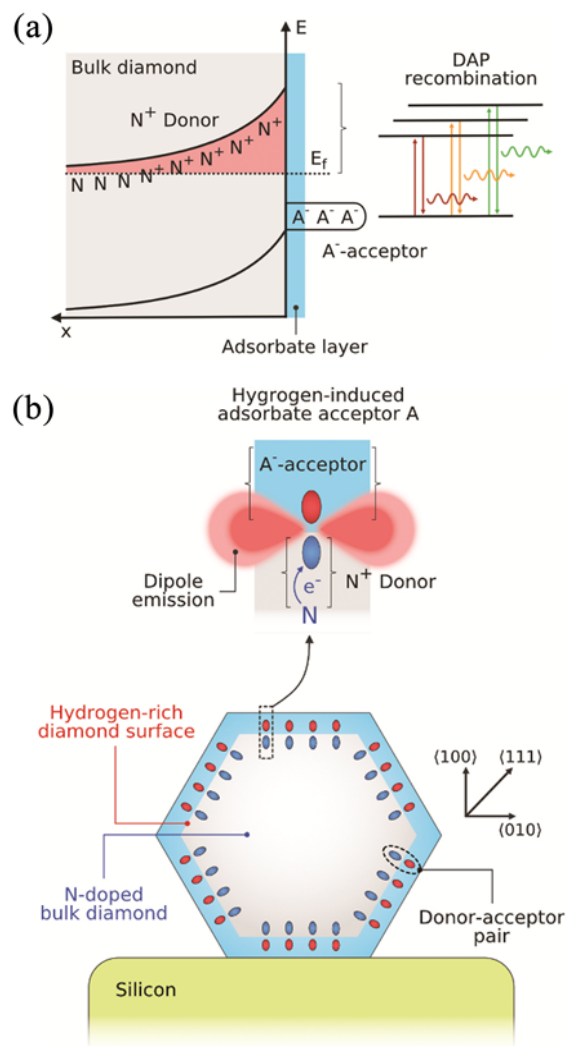

*Figure 5. (a) Sketch of space charge distribution and band bending profiles near the surface of N-doped hydrogenated diamond. (b) Schematic representation of the location and orientation of DAP dipoles near the surface of N-doped hydrogenated diamond crystallite.*

**Conclusion**

In summary, we have established a connection between the mysterious NFLs observed in hydrogenated nanodiamond in a wide spectral range (500 – 800 nm / 2.48-1.55 eV) with nitrogen impurity forming donor centers in the ND volume and a 2D layer of surface acceptors resulting from the transfer doping effect. In particular, the connection between NFLs and surface acceptors is confirmed by an increase in the number of lines with an increase in the surface/volume ratio in the ND cluster under study, and vice versa, by a decrease in the number of lines with a decrease in the number of surface acceptors through thermal oxidation of the ND surface. The connection between NFL and donor-like nitrogen is confirmed by a significant increase in the number of lines upon intensive doping of NDs with nitrogen. Thus, these results with a high degree of confidence confirm our hypothesis that new lines in the spectra of hydrogenated HPHT diamonds arise at radiative recombination of DAPs, formed under the interaction of nitrogen donors and surface acceptors.

We found that discrete NFLs are associated predominantly with single DAPs, which are single-photon emitters. The characteristic linewidth of such emitters at room temperature is 0.5-2 nm (1.5 -6 meV), the lifetime of their excited state is 1-2 ns, the photon emission rate can exceed 1 million counts/s. A specific behavior of individual lines is observed as T decreases to 10 K. The lines first shift towards higher energies, and then in the range of 100-10 K their energy slightly (~50 pm) decreases. In addition, at T less than 100 K, the lines split into 2 or more components. The only problem of the DAPs was found to be temporary instability of their fluorescence, characteristic of emitters unprotected from environment. Finding ways to stabilize the DAP radiation in hydrogen-terminated nanodiamonds opens new horizons for room-temperature quantum photonics.

## Methods

### Nanodiamonds Synthesis and Sample Preparation

Three types of the ND samples were synthesized using the high pressure technique [25]. The first type is undoped ND powder with particle sizes ranging from 100 nm to 1 μm, prepared from a mixture of adamantane $C_{10}H_{16}$ (Sigma Aldrich, >99%) and octafluoroaphthalene ($C_{10}F_8$, Alfa Aesar, 96% purity) in a weight ratio of 1:1. The second type is a nitrogen doped ND powder of the same particle size as the first type, prepared from a mixture of adamantane and adamantane carbonitrile $C_{11}H_{15}N$ (>97%, Sigma Aldrich) in a weight ratio of 50:1. Both types of samples were synthesized at about 1600÷1700°C. The third type of samples are nitrogen-doped ND powders synthesized from a mixture of adamantane $C_{10}H_{16}$ (Sigma Aldrich, >99%) and detonation NDs (5 nm in size, Adamas Nanotechnologies, Inc.) in a weight ratio of 10:1. NDs of this type synthesized at two temperatures, approximately 1600÷1700°C and 1300÷1400°C, consist of particles of 100 nm ÷ 1 μm and <50 nm size, respectively.

The initial chemical reagents were manually mixed in an agate mortar, pressed into a cylindrical pellet and placed in a graphite capsule. All three types of nanodiamonds were synthesized at a pressure of 7.5-8 GPa. The exposure time at temperature was 7 to 20 seconds. The mass of the synthesized diamond powder was ~10 mg. A probe was taken from the central part of the samples and dispersed in ethanol. A small volume of the resulting suspension was then dropped onto a silicon substrate and dried. As a result, a layer of individual and/or clustered diamond particles was formed on the Si surface.

### Electron Microscopy

The crystal size and morphology were studied by a SEM Jeol 7001F and scanning-tunneling TEM (STEM) Talos F200S (FEI, USA), operated at 200 kV. The sample preparation for TEM was performed in a conventional way - depositing of the sample from the ethanol suspension on a lacey carbon supported copper grids (TED Pella).

### Fluorescence Study

Room temperature PL spectra were measured with a LabRam HR800 spectrometer in a confocal configuration with a spatial resolution of 1 μm. The PL was excited with a 473 nm, 532 nm and 660 nm diode lasers and collected in backscattering geometry with the microscope objective (Olympus, magnification 100, the numerical aperture NA 0.90). The laser power reaching NDs was 0.1–4 mW.

For NFL characterization at low temperature two different homebuilt setups were used.

The first one is equipped with 515 nm (CW and pulsed mode, Omicron QuixX® 515-80 PS) and 660 nm (CW, Thorlabs L660P120) diode lasers with the pump power from 0.1 to 1 mW. In the cryogenic experiments, the sample was mounted in a chamber of a close- loop cryostat (atto DRY800, Attocube). The temperature of the sample holder was stabilized at around 10 K. The cryostat was integrated with a home-built confocal laser-scanning microscope. The beam of the pump laser was focused on the sample by cryo-compatible microscope objective lens (NA=0,82, Attocube). The position of the pump beam on the sample was scanned using a galvo mirror (FSM 300, Newport). The signal was collected into a single-mode fiber (SM600, Thorlabs) and then sent either to a single-photon avalanche photodiode (SPAD, SPCM-AQRH-15, Perkin Elmer) or a spectrometer (Princeton Instruments IsoPlane 160, max resolution 0.13 nm, 1200 gr/mm).

To determine the purity of single photon emission, the Hanbury Brown–Twiss correlation scheme was used based on the 50:50 fiber beamsplitter (Thorlabs). For the time-resolved kinetic PL studies the 515 nm laser was set to the pulsed mode (110 ps pulse width, 2 MHz pulse rate). Both for the

$g^{(2)}$ and lifetime measurements the photocounts were processed by the time-to-digital converter (Qutools quTAU, 81 ps resolution).

In the second setup a sample was mounted in a helium cryostat (Janis ST-500) cooled from 300 K to near 10 K. Diode CW 532 nm laser radiation was focused on the sample using a x100 objective with a numerical aperture of 0.7 (Mitutoyo Plan Apo NIR). PL spectra were measured at 0.1 - 0.8 mW laser power reaching the ND. Incident radiation was focused in 2-3 µm spot on the sample. The PL was collected with the same objective and analyzed using a grating spectrometer (Princeton Instruments SpectraPro HRS-500 with 1800 gr/mm) combined with a cooled CCD camera (Princeton Instruments PyLoN). For polarization analysis a halfwave plate and a film polarizer installed in the detection channel were used.

**Sample annealing**

NDs were annealed in the air in a LinKam TS1500 chamber at a constant temperature of 620°C for 30 min. The heating rate was 120°C/min.

**Acknowledgement**

D.K. and V.L. acknowledge the funding support from Agency for Science, Technology and Research (C230917005). A.L.C. acknowledged Bridge POC grant by SNF and Innosuisse (Grant No. 40B1-0_205841 Nano-IonEX membrane).